\input harvmac.tex

\def\nup#1({Nucl.\ Phys.\ $\bf {B#1}$\ (}
\def\l{\lambda}
\noblackbox
\Title{\vbox{
\hbox{HUTP-98/A069}
\hbox{\tt hep-th/9809187}
}}{M-Theory and Topological Strings--I}
\bigskip
\centerline{Rajesh Gopakumar \foot{gopakumr@tomonaga.harvard.edu}and  Cumrun
Vafa \foot{vafa@string.harvard.edu}}
\bigskip
\centerline{Lyman Laboratory of Physics}
\centerline{Harvard University}
\centerline{Cambridge, MA 02138}

\vskip .3in
The $R^2 F^{2g-2}$ terms of Type IIA strings on Calabi-Yau 3-folds,
which are given by the corresponding
topological string amplitudes
(a worldsheet instanton sum for all genera),
 are shown to have a simple M-theory interpretation.
 In particular, a Schwinger
one-loop computation in M-theory with wrapped M2 branes
and Kaluza-Klein
modes going around the loop reproduces the all genus string
contributions from constant maps and worldsheet instanton
corrections. In the simplest case of an isolated M2 brane
with the topology of the sphere, we obtain the contributions of
small worldsheet
instantons (sphere ``bubblings'') which extends the results known or
conjectured for low genera.  Surprisingly, the
't Hooft expansion of large $N$
Chern-Simons theory on $S^3$
can also be used in a novel way to compute
these gravitational terms at least in special cases.

\Date{September 1998}

\lref\strom{A. Strominger, ``Massless Black Holes and Conifolds in String
Theory'', \nup451 ,(1995), 96 .}

\lref\pol{J. Polchinski, ``Dirichlet-Branes and Ramond-Ramond Charges'',
Phys.Rev.Lett. {\bf 75}(1995), 4724.}

\lref\witym{E. Witten, ``Bound States Of Strings And $p$-Branes'',
\nup460 (1996), 335 .}

\lref\strovaf{A. Strominger, C. Vafa, ``Microscopic Origin of the
Bekenstein-Hawking Entropy'', Phys.Lett. {\bf B379} (1996), 99 .}

\lref\narcon{I.Antoniadis, E.Gava, K.S.Narain, T.R.Taylor,
``N=2 Type II- Heterotic duality and Higher derivative
F-terms'', \nup455 (1995), 109.}

\lref\jose{J. Morales, M. Serone,
``Higher Derivative F-terms in N=2 Strings'',
\nup481 (1996), 389}

\lref\naret{I. Antoniadis, E. Gava, K.S. Narain and T.R. Taylor,
``Topological Amplitudes in String Theory'', \nup413 (1994), 162.}

\lref\BFSS{T. Banks, W. Fischler, S.H. Shenker, L. Susskind,
``M Theory As A Matrix Model: A Conjecture'', Phys.Rev. D55 (1997), 5112 .}

\lref\DVV{R. Dijkgraaf, E. Verlinde, H. Verlinde,
``Matrix String theory'', \nup500 (1997), 43 .}

\lref\BS{T. Banks, N. Seiberg, ``Strings from Matrices'',
\nup497 (1997), 41 .}

\lref\juan{J. Maldacena, ``The Large N Limit of Superconformal Field Theories
and
Supergravity'', hep-th/9711200 .}

\lref\hora{P. Horava, ``M-Theory as a Holographic Field Theory'',
hep-th/9712130 .}

\lref\ghovaf{D. Ghoshal, C. Vafa, ``c=1 String as the Topological Theory of
the Conifold'' , \nup453 (1995), 121 .}

\lref\joao{A. Jevicki, M. Mihailescu, J. P. Nunes,
``Large N WZW Field Theory Of N=2 Strings'',hep-th/9706223 .}

\lref\mart{M. O'Loughlin, ``Chern-Simons from Dirichlet 2-brane instantons'',
Phys.Lett. {\bf B385} (1996) 103 .}

\lref\emil{E. Martinec, ``M-theory and N=2 Strings'',
hep-th/9710122 .}

\lref\witstr{E. Witten -- top.strings.}

\lref\witcs{E. Witten, ``Chern-Simons Gauge Theory As A String Theory'',
hep-th/9207094.}

\lref\vfr{V. Balasubramanian, R. Gopakumar, F. Larsen,
`` Gauge Theory, Geometry and the Large N Limit'', hep-th/9712077 .}

\lref\bcovi{M.Bershadsky, S.Cecotti, H.Ooguri, C.Vafa , ``Holomorphic
Anomalies in Topological Field Theories, \nup405 (1993), 279 .}

\lref\bev{N. Berkovits, C. Vafa,
``N=4 Topological Strings'', \nup433 (1995), 123 .}

\lref\besi{N. Berkovits, W. Siegel ,
``Superspace Effective Actions for 4D Compactifications of
Heterotic and Type II Superstrings'', \nup462 (1996), 213 .}

\lref\ova{ H. Ooguri, C. Vafa, ``Geometry of $N=2$ Strings'',
\nup361 (1991), 469 .}

\lref\ovhe{ H. Ooguri, C. Vafa, ``$N=2$ Heterotic Strings'',
\nup367 (1991), 83 .}

\lref\dv{J. Distler, C. Vafa, ``A Critical Matrix Model at c=1'', Mod. Phys.
Lett. A6, (1991), 259 .}

\lref\wit{E. Witten, ``Quantum Field theory and the Jones Polynomial'',
Commun.Math.Phys. {\bf 121},(1989) 351 .}

\lref\grossk{D. Gross, I. Klebanov, ``One Dimensional String theory on a
Circle'',
\nup344 (1990), 475 .}

\lref\kazet{V.A.Fateev, V.A.Kazakov, P.B.Wiegmann,
``Principal Chiral Field at Large N'', \nup424 (1994), 505 .}

\lref\mva{S. Mukhi, C. Vafa, ``Two dimensional black-hole as a topological
coset model of c=1 string
   theory'', \nup407 (1993) 667 .}

\lref\gepwit{D. Gepner, E. Witten, ``String Theory on Group Manifolds'',
\nup278 (1986), 493 .}

\lref\lisa{L. Jeffrey, ``Chern-Simons theory on Lens Spaces and Torus Bundles,
and the Semi-classical
Approximation'', Commun.Math.Phys.147, (1992), 563 .}

\lref\roza{L. Rozansky, Commun.Math.Phys. 175 , 275 .}

\lref\wit{E. Witten, ``Quantum Field Theory and the Jones Polynomial'',
Commun.Math.Phys.121, (1989), 351 .}

\lref\peri{V. Periwal, `` Topological Closed-string Interpretation of
Chern-Simons Theory'', Phys. Rev. Lett. {\bf 71} (1993), 1295.}

\lref\ovan{H. Ooguri, C. Vafa, ``All Loop N=2 String Amplitudes'',
\nup451 (1995) 121 .}

\lref\vafcon{C. Vafa, ``A Stringy Test of the Fate of the Conifold''
\nup447 (1995), 252 .}

\lref\wigrr{E. Witten, `` Ground Ring Of Two Dimensional String Theory'',
\nup373 (1992), 187\semi E. Witten and B. Zwiebach, Nucl. Phys.
{\bf B377} (1992) 55.}

\lref\polk{I. Klebanov, A. Polyakov, ``Interaction of Discrete States in
Two-Dimensional String Theory'',
Mod. Phys. Lett. A6 (1991), 3273 .}

\lref\modp{R. Dijkgraaf, G. Moore, R. Plesser, ``The partition function of 2d
string theory'',
\nup394 (1993), 356 .}

\lref\BSV{M. Bershadsky, V. Sadov, C. Vafa, ``D-Strings on D-Manifolds'',
\nup463 (1996), 398 .}

\lref\gopvaf{R. Gopakumar, C. Vafa, ``Branes and Fundamental Groups'',
hep-th/9712048.}

\lref\iz{C. Itzykson, J. Zuber, ``Quantum field theory'', Addison Wesley
Publishing, 1980.}

\lref\dilbas{D. Jatkar, B. Peeters, ``String Theory near a Conifold
Singularity'',
Phys.Lett. B362 (1995), 73 .}

\lref\esalb{S. Kachru, E. Silverstein, A. Lawrence, ``On the Matrix Description
of Calabi-Yau Compactifications''
 hep-th/9712223 .}

\lref\bcovii{M. Bershadsky, S. Cecotti,H. Ooguri, C. Vafa ,
``Kodaira-Spencer Theory of Gravity and Exact Results for
Quantum String Amplitudes'', Comm. Math. Phys. {\bf 165} (1994), 311 .}

\lref\twobcs{R. Gopakumar, C.Vafa, ``Topological Gravity as a Large N
Gauge theory'', Adv. Theor.Math.Phys. {\bf 2} (1998), 413}

\newsec{Introduction}

The relationship between eleven dimensional M-theory and
type IIA string theory has been a fruitful source of insight
since its discovery. Specifically, since fundamental
strings and D-branes of type IIA are on a more
unified footing in eleven dimensions, we can often say something
non-perturbative about string theory from simple considerations in
M-theory.

In this paper we concentrate on Type IIA strings on Calabi-Yau
threefolds.
The full IIA theory compactified on a Calabi-Yau threefold
contains terms in the low energy action which are
captured by the simpler $N=2$ topological string
theory.  In this paper, we will
reinterpret the worldsheet instanton sum of the latter in
terms of a one loop computation in M-theory. This generalizes
some of the observations made in \ref\nekla{A. Lawrence and N. Nekrasov,
``Instanton sums and five-dimensional gauge theories,''
Nucl. Phys. {\bf B513} (1998) 239.}\
about obtaining the genus 0 topological amplitudes (the prepotential) from
M-theory on a Calabi-Yau times a circle.  It is also similar
in spirit to the observations in \ref\gregu{M.B. Green, M. Gutperle, P.
Vanhove,``One loop in eleven dimensions,''
Phys. Lett. {\bf B409} (1997) 177.}\
relating $R^4$ amplitudes of type IIA to Kaluza-Klein contributions
in M-theory.

The topological string
theory computes certain  F-terms in the low energy four dimensional
theory. We briefly review this in Sec.2.
Since some of these quantities are coupling
constant independent, they can equally well be computed
at strong coupling. M-theory provides a simple, physical way of doing this.
For a large
Calabi-Yau, in this limit, the lightest relevant objects are Kaluza-Klein
modes (D0 branes). Integrating them out in a Schwinger type computation gives
us precisely the leading contribution at every genus from constant maps
(the whole worldsheet mapped to a point)
in the topological string
theory. The next to leading contributions come from M2 branes
wrapped on Riemann surfaces in the Calabi-Yau. In this paper we will
treat the simplest case when the surface is an isolated
$S^2$ in the Calabi-Yau.\foot{The higher genus M2 branes (as well as
families of them) will be treated in a subsequent paper
\ref\nextp{R. Gopakumar and C. Vafa, ``M-theory and Topological
Strings--II,'' to appear.}.}
There can be additional
Kaluza-Klein modes here too, corresponding to bound states of $D2-D0$
branes. Again, we can sum their one loop contributions exactly.
What we obtain is usually interpreted in the string language as
contributions from small worldsheet instantons. The one loop Schwinger
computation has non-perturbative information as well,
allowing us to compute
non-perturbative corrections to topological string amplitudes.
In fact, we can add together the perturbative and non-perturbative
contributions and write the full
partition function for the topological string
in a suggestive way.
All this will be the
subject of Sec.3.
In Sec.4, we consider open topological strings on $T^*S^3$ with dirichlet
boundary conditions on the $S^3$.  We can compute the open topological
string amplitudes using the large $N$ Chern-Simons theory,
to which it is equivalent
\witcs . By extrapolating the expressions for $h$ boundaries, to the case
$h=0$,
we surprisingly recover the leading topological {\it closed} string amplitude.
This is a novel ``$h\rightarrow 0$'' way of recovering a gravity
result from (open string) gauge theory. The open string expansion itself
seems to also reproduce the subleading contribution from 2-branes on an
$S^2$, though we do not have a deep understanding of this aspect of it.

After we had obtained the results of the present paper, we received a paper
\ref\mor{M. Marino and G. Moore, ``Counting higher genus curves in a Calabi-Yau
manifold,'' hep-th/9808131.}\ which has some overlap with ours, but uses
very different arguments.

\newsec{Topological Strings and What They Compute}
Consider type IIA strings compactified on a Calabi-Yau
threefold to 4 dimensions. We obtain an $N=2$ supersymmetric
theory, with vector and hypermultiplet fields. The scalars
in these multiplets are moduli belonging to the Kahler moduli space
and (a jacobian variety over) the complex moduli space respectively.
For type IIB strings it is much the same, except that the role of Kahler
and complex moduli are exchanged. There are a special class of
F-terms which do not mix the two types of moduli.

For example, in the IIA theory
the F-terms of the low energy theory
 involving only Kahler moduli are of the form
${\cal F}_g R_+^2 F_+^{2g-2}$. Here, $+$ denotes the self-dual parts of the
curvature, $R$ denotes the Riemann tensor and $F$ the graviphoton
field strength. (We will be considering the euclidean effective action.)
\foot{The contractions of indices here are determined
from the fact that this term comes from $ \int d^4\theta W^{2g}$ in
superspace. Here $W$ is a Weyl superfield, $W_{\mu\nu}=F^+_{\mu\nu}
-R^+_{\mu\nu\l\rho}\theta\sigma^{\l\rho}\theta +....$}
The coefficient, ${\cal F}_g$ is purely a function of Kahler
moduli. Rather remarkably, the ${\cal F}_g$ can be computed
using a much simpler string theory, an $N=2$ topological string,
with the Calabi-Yau threefold as its target space \bcovii \naret .
In fact, ${\cal F}_g$ is the partition function of the perturbative A-model
topological closed string theory at genus $g$.

One can show that the above superpotential terms receive contributions
only from genus $g$ amplitudes in the physical string theory. Moreover,
there are no non-perturbative string corrections.
(The dilaton lies in a hypermultiplet and cannot mix with the vector
multiplet moduli -- the same statement
would therefore not be true for superpotential terms involving
hypermultiplets).
Thus, the partition function of topological strings computes certain
exact quantities in the full, physical string theory.

In the limit of large volume (radius) of the Calabi-Yau three-fold, the
${\cal F}_g$ admits a purely topological interpretation: It is roughly given by
the worldsheet instanton sum
\eqn\parti{{\cal F}_g=\sum_{C_g}exp(-A_C).}
The sum is over Riemann surfaces (holomorphic curves) $C_g$ of genus $g$
embedded in the Calabi-Yau
threefold -- the target space images of the worldsheet.
And $A_C$ denotes the complex area (including the imaginary
contribution from the B-field) of $C$.  This, however, is not
the full story: Sometimes, there are whole
families of holomorphic curves embedded in the three-fold. The sum is now
really an integral and one has to weight it appropriately. The integral
is over an appropriate moduli space
and the relevant object that enters as the weight turns out to be
an appropriate cohomology class over
this moduli space.
In addition, there can be further contributions from
degenerate genus $g$ curves. In fact, the leading large radius
contribution to \parti\ comes from constant maps where the whole
genus $g$ worldsheet is mapped to a point in the Calabi-Yau. (This is
the leading contribution
in the large volume limit, since
it is not suppressed by the area exponent.)
In this case, the relevant moduli space is
the product,
${\cal M}_g$ times the Calabi-Yau threefold $K$ itself (corresponding to
the choice of the point
in the target space). ${\cal M}_g$ is the familiar moduli space
of all Riemann surfaces of genus $g$.
Then the appropriately weighted contribution to ${\cal F}_g$ from constant maps
(which, as we
remarked, is the leading contribution for large volumes),
was found  by \bcovii\ to be
\eqn\raje{{\cal F}_g = {1\over 2}\chi_{K} \int_{{\cal M}_g} c_{g-1}^3+
O(exp(-A))}
Here $\chi_{K}$ denotes the Euler characteristic of $K$ and
$c_{g-1}$ denotes the $(g-1)$-th chern class of the Hodge bundle
over ${\cal M}_g$ (The Hodge bundle is the $g$-dimensional holomorphic
vector bundle over ${\cal M}_g$
locally spanned by the  $g$ holomorphic 1-forms on
the genus $g$
Riemann surface $g>1$). $\int_{{\cal M}_g} c_{g-1}^3$ has only very
recently
been computed by mathematicians to be
(\ref\math{C. Faber and R. Pandharipande, ``Hodge integrals and
Gromov-Witten theory,''
in preparation.})
\eqn\cgeqn{\int_{{\cal M}_g} c_{g-1}^3={B_g\over 2g(2g-2)}{B_{g-1}\over
(2g-2)!}=(-1)^{g-1}\chi_g{2\zeta(2g-2)\over (2\pi)^{2g-2}}.}
Here $\chi_g=(-1)^{g-1} {B_g\over 2g(2g-2)}$ is the euler characteristic of
${\cal M}_g$. ($B_g$ are the Bernoulli numbers taken here to be all
positive.)
The second way of writing this expression in
terms of $\chi_g$ and the Riemann zeta function
$\zeta$ was motivated by the physical derivation we shall
present in this paper for this leading contribution.

There are further kinds of degenerate
contributions, known as ``bubblings''.
The simplest instance of this is when all of a genus $g$ worldsheet,
except the infintesimal neighborhood of a single point,
is mapped to a point on an $S^2$ in the Calabi-Yau, while this neighborhood
itself
wraps the rest of the $S^2$. From the two-dimensional worldsheet point of
view, this is what is sometimes called a ``small'' instanton.
The general case similarly involves
a genus $g$ curve which has a lower genus
part which is mapped non-trivially to a genus $g^{\prime}$
($g > g^{\prime}$) curve in the Calabi-Yau,
while the rest of the surface is mapped to a point.
These bubblings were first encountered in computing ${\cal F}_1$
which has a contribution from a sphere bubbling off of a torus
\bcovi\
and was explained by Katz (see the appendix of \bcovi\ by
Katz).
Subsequently in \bcovii,
the contribution of genus 0 and genus 1 bubbling off of a genus 2 surface was
conjectured (the genus 1 contribution was actually conjectured to be zero).
Since the computation of higher genus ${\cal F}_g$ for $g>2$
has not been carried out, the structure of degenerate
contributions has not even been conjectured.
We will see that the one loop amplitude in M-theory involving
Kaluza-Klein states of massless fields as well as Kaluza-Klein
modes of wrapped M2 branes reproduces and extends these
topological string amplitudes to all perturbative orders as well as imply
further non-perturbative corrections.   In other words, all these
contributions to the $R^2F^{2g-2}$ amplitudes in the IIA theory can be
alternatively viewed as
coming from integrating out wrapped brane states.

\newsec{M-theory and IIA amplitudes}

M-theory compactified on a circle is conjectured to be equivalent to
type IIA strings \ref\townsWit{P. Townsend, ``The eleven dimensional
supermembrane revisited,''
Phys. Lett. {\bf B350} (1995)
184\semi E. Witten, ``String theory dynamics in various dimensions,''
Nucl. Phys. {\bf B443} (1995) 85.}, where Kaluza-Klein
modes of the circle become equivalent to bound states of D0 branes.
In the limit of a small $S^1$ in the eleventh dimension, we recover
the perturbative regime of type IIA strings. Again, the M2 branes of
M-theory reduce to fundamental strings or
D2 branes of type IIA
depending on whether or not they wrap around the $S^1$.
Thus, M-theory on a large radius Calabi-Yau threefold times a small $S^1$
is equivalent
to perturbative type IIA strings on the same large Calabi-Yau.
In particular, it is in this limit that
the F-terms of Sec. 2 are given by a
worldsheet instanton sum.  Now consider
the limit where the circle grows, corresponding to increasing
the type IIA coupling constant.  Since the superpotential
computations (depending on Kahler moduli) are exact they
are unmodified in this limit.  But now, we have the perspective of M-theory
and we can ask how the same computation would look from this vantage point.
This is the basic strategy we follow in the next subsection.

\subsec{M-theory Computation of Topological String Amplitudes}
In the limit of large string coupling of the IIA theory on a big Calabi-Yau
threefold, the light objects from the IIA viewpoint
are D0 branes and their bound states. As mentioned above, these are
the Kaluza-Klein modes of massless fields of M-theory on the Calabi-Yau
threefold.

First, let us determine from the type IIA perspective
what the precise number of $D0$ brane bound states are, in this geometry.
On $R^{10}$ we have one bound state of $n$ D0-branes for each $n$.
The D0-branes are point like objects on the Calabi-Yau
threefold times $R^3$. Therefore the quantum mechanical
supersymmetric ground states
are in one to one correspondence to the compact cohomology elements of the
Calabi-Yau (times a hypermultiplet). Thus for each cohomology element and each
integer $n$ we obtain a state in type IIA with mass ${2\pi n\over \l}$
where $\lambda$ is the string coupling. (It will be convenient in the
following
to work in
units where $\alpha^{\prime}={1\over 4\pi^2}$.) These are exactly the same
as the $S^1$ Kaluza-Klein massless modes of M-theory
compactified on the Calabi-Yau.  Since they are charged under
the $U(1)$ graviphoton, they will have a one loop contribution to
$R^2 F^{2g-2}$, which we will now compute.

Let us consider the contribution of one 4 dimensional $N=2$
hypermultiplet to
${\cal F}_g$.  Let $Z$ denote the central terms in the supersymmetry
algebra for this hypermultiplet, where the mass $m=|Z|$.  Then following
the argument in \ghovaf\ and the direct ``Schwinger-type'' computation
of \narcon\ one deduces that the contribution
of this hypermultiplet to ${\cal F}_g$ ($g>1$ for most of
the expressions in this section, though most statements
hold more generally after an appropriate continuation) is
\eqn\conh{{\cal F}_g= -\chi_g Z^{2-2g}}
where $\chi_g=\chi({\cal M}_g)$ as before.
%
%
After taking account of all the multiplets
for a fixed Kaluza-Klein momentum around the circle, the net contribution
turns out to be $h^{2,1}(K)-h^{1,1}(K)={-\chi (K)/2}$ times the contribution of
a single hypermultiplet\foot{This involves
a somewhat subtle contribution from various
massive vector, hyper and spin 2 multiplets, which
will be explained in more detail in \nextp .}.  Using \conh\ , with $Z={2\pi
in\over \l}$ and summing over the contribution
of all Kaluza-Klein modes we find (for the coefficient of $\l^{2g-2}$)
$${\cal F}_g=\chi_g {\chi_K \over 2}\sum_{n\in {\bf Z}, n\neq 0} (2\pi
in)^{2-2g}$$
Comparing this with the expected large radius behaviour of topological
amplitude in equation \raje\ gives us
$$\int_{{\cal M}_g}c_{g-1}^3 = (-1)^{g-1}\chi_g{2\zeta(2g-2)\over
(2\pi)^{2g-2}}$$
precisely agreeing with the recent mathematical results \cgeqn\ .

Actually, we can say more from M-theory. The Schwinger computation is
famous for its prediction of non-perturbative pair creation. If we consider
a constant self-dual graviphoton field, then in addition to the
contribution to $R_+^2$ from the terms proportional to $F^{2g-2}$, there
will be those from terms behaving like $e^{-{1\over F}}$.
The expression derived by Schwinger for the one-loop determinant
captures both perturbative and non-perturbative parts. In our context, for a
particle of BPS charge $Z$ in a constant self-dual field $F$, the
expression is
(see for example \iz\ )
\eqn\fmu{{\cal F}(Z)=\int_{\epsilon}^{\infty} {ds\over s^3}
({s/2 \over \sinh s/2})^2 e^{-{sZ\over F}}}
(where $\epsilon \rightarrow 0^+$ is a cutoff).  This has a
perturbative expansion that agrees with \conh\
$${\cal F}(Z)=\sum_{g}{\cal F}_gF^{2g-2}+O(e^{-{Z\over F}}).$$
In our case with $Z={2\pi in\over \l}$, we can easily carry out the sum
over $n\in {\bf Z}, n\neq 0$, using $\sum_{n=-\infty}^{\infty}e^{in\theta}
=2\pi\sum_{m=-\infty}^{\infty}\delta(\theta-2\pi m)$. Also taking into
account the factor of ${\chi_K\over 2}$, we have the complete zero brane
contribution to be
$${\cal F}(Z)={\chi_K\over 8}\sum_{m=1}^{\infty}
{1\over m}{1\over (sinh{m\l F\over 2})^2},$$
after dropping an irrelevant field independent term.
This sum can be more suggestively written after expanding out the denominator
and carrying out the sum over $m$, as
$${\cal F}(Z)=-{\chi_K\over 2}\sum_{n=1}^{\infty}n\ln(1-q^n) ; \qquad q=e^{-\l
F}.$$

Let us now consider the contribution of isolated M2 branes with the
topology of $S^2$.  This gives a hupermultiplet in 5 dimensions.
Let $A$ denote the area of the sphere.  Consider
the Kaluza-Klein modes of these states (which in the type
IIA description correspond to bound states of a single D2 brane with a number
of D0 branes).  Then the central term in the susy algebra
of the state with $n$ units of Kaluza-Klein momentum is given by
$Z=2\pi{(A+ i n)\over \l}$ and so using \conh\ we get its contribution to
${\cal F}_g$ to be
\eqn\vvv{\eqalign{{\chi_g \over (2\pi)^{2g-2}}\sum_{n\in {\bf Z}} (A+
in)^{(2-2g)}=&
{-\chi_g\over (2\pi)^{2g-2}(2g-3)!}({d\over dA})^{2g-2}ln(1-e^{-2\pi A})\cr =&
\chi_g{1\over
(2g-3)!}\sum_n n^{2g-3}e^{-2\pi nA}.}}
(Here we have used the product formula ${sinh(\pi x)\over \pi
x}=\prod_{n=1}(1+{x^2\over n^2})$.)
This agrees, for $g=1,2$, with the contribution from degenerate genus zero
instantons \bcovi\bcovii\ \foot{For the genus 0 and 1
the above expression should be interpreted
in the regularized form $\chi_g/
(2g-3)!=1,{-1/12}$ respectively.}
and generalizes it to spherical
bubbling contributions to an arbitrary genus topological amplitude\foot{One
may wonder about the existence of bound states
of multiple $D2$ branes wrapped over $S^2$, but using the methods
in \ref\bsv{M. Bershadsky, V. Sadov and C. Vafa, ``D-branes and
topological field theories,''
Nucl. Phys. {\bf B463} (1996)
420.}\ one can show that there are none.  This is {\it not} the case
for multiple $D2$ branes wrapped on higher genus surfaces and
they do lead to contributions to topological amplitudes \nextp .}.

Once again, the Schwinger computation gives us non-perturbative
information, now about pair creation of bound states of 2-branes with
0-branes in a graviphoton field.
With $Z=2\pi{(A+in)\over \l}$, we can go over steps similar
to before to obtain
$${\cal F}(Z)={1\over 4}\sum_{m=1}^{\infty}{e^{-2\pi mA}\over m}{1\over
(sinh{m\l
F\over 2})^2}$$
or alternatively
$${\cal F}(Z)=-\sum_{n=1}^{\infty}n\ln(1-zq^n) ; \qquad z= e^{-2\pi A}.$$

Thus the combined contributions to the free energy can be written as
$${\cal F}_{tot}(Z)=-\sum_{n=1,[n_i]}^{\infty}n\ln[(1-q^n)^{\chi_K\over
2}(1-q^n\prod_i z_i^{n_i})^{d_{[n_i]}}],$$
where we are including the contribution of all isolated genus 0
$M2$ branes, assuming there are $d_{[n_i]}$ of them in a homology class
labelled by $[n_i]$ and $-{\rm log}z_i$ denotes the
area of the corresponding class.  These formulae are
very much reminiscent of some observations in
\ref\harmor{J. Harvey and G. Moore, ``Algebras, BPS States, and Strings,''
Nucl.Phys. {\bf B463} (1996) 315.}.

Alternatively, with $\l F\rightarrow \l$, this can be thought of as
a portion
of the non-perturbative topological string partition function. We also
recall that a similar
Schwinger computation for wrapped D3 branes had yielded non-perturbative
contributions to the full B-model topological string in the vicinity
of a conifold \twobcs . In that case, a connection with Chern-Simons theory had
also been
made. In the next section we will see that a connection to Chern-Simons
theory exists for our A-model closed string amplitudes as well.

\newsec{Chern-Simons theory in the 't Hooft Expansion}

Chern-Simons theory enjoys the advantage of being an exactly
solvable gauge theory \wit . Therefore its large N expansion offers
an instance where one can study the correspondence to string theory
explicitly. Periwal \peri\ has studied the $N\rightarrow
\infty$ limit of this theory on $S^3$. In \twobcs\ it was shown that the free
energy of the gauge
theory in the $N=\infty$ limit matched that of the {\it B-model}
topological closed string in the vicinity of the conifold (and some
generalizations). The gauge theory,
remarkably enough, seems to capture both
the perturbative and non-perturbative features of the closed string
theory\foot{
For another possible connection between Chern-Simons theory and type IIA string
theory
instantons see \mart .}.

But here, we will be concerned with the alternative interpretation
by Witten of Chern-Simons theory as an {\it open} string theory. This goes
as follows. In any large N gauge theory with adjoint fields only, the
't Hooft expansion for, say, the free energy can be written as
\eqn\tHooft{F=\sum_{g=0,h=1}C_{g,h}N^{h}\kappa^{2g-2+h}
=\sum_{g=0,h=1}C_{g,h}N^{2-2g}\lambda^{2g-2+h}}
Here $\kappa$ is the ordinary gauge coupling while $\lambda=\kappa N$ is the
't Hooft coupling which is held fixed in the large $N$ limit. As usual $g$ is
the
genus of the Riemann surface built out of double index lines. While $h$ is
the number of closed index loops (or faces of the triangulated Riemann
surface) which accounts for the factor of $N^h$. It is not to be confused
with the holes that appear in the presence of matter in the fundamental.

What Witten argued in the particular case of Chern-Simons theory on a
three manifold $M$ was that $C_{g,h}=-Z_{g,h}$, where $Z_{g,h}$ is
the partition function of the
A-model topological  open string theory at genus $g$ with $h$ boundaries
on the 6-dimensional target space $T^*M$.
Thus the 't Hooft expansion tells us about the
complete perturbative expansion of an open string theory.

In the case when $M=S^3$, the coefficients $C_{g,h}$ for the $SU(N)$ level
$k$
theory are not too difficult
to compute explicitly (see the appendix). The main subtlety to keep in mind
is the fact that
the bare Chern-Simons coupling undergoes a finite quantum renormalization.
Therefore, the 't Hooft expansion is performed with $\lambda={2\pi N\over
k+N}$.
(To connect the 't Hooft expansion in the appendix with the open string
expansion,
we must take $\l \rightarrow i\l$, due to the $i$ that is present in the
definition of the Chern-Simons action. This introduces the extra factor of
$(-1)^{g-1+p}$ in the answer below.)
The final answer then is fairly simple:
\eqn\cgh{C_{g,2p}=(-1)^{g-1+p}\chi_{g,2p}{\zeta(2g-2+2p)\over (2\pi)^{2g-2+2p}}
;\quad C_{g,2p+1}=0}
where
\eqn\euler{\chi_{g,h}= (-1)^h{\eqalign{\pmatrix{2g-3+h\cr h}}}\chi_g =
(-1)^h{\eqalign{\pmatrix{2g-3+h\cr h}}}(-1)^{g-1} {B_g\over 2g(2g-2)}}
is the Euler characteristic of the moduli space of Riemann surfaces with
genus $g$ and $h$ punctures (see \dv\ for a connection to the $c=1$ string).
Note that the
't Hooft expansion starts off with at least one hole (actually
in this case all the amplitudes with an odd number of holes vanish
and the first term is one with 2 holes).  Let us consider
$C_{g,2p}$ as an analytic function in $h=2p$.  The case
with $h=0$ would have corresponded to {\it closed} topological
strings.  In this limit we should get
$$Z_{g,h\rightarrow 0} \rightarrow  {\chi(T^* S^3)\over 2}\int_{{\cal M}_g}
c_{g-1}^3.$$
If we take $\chi(T^* S^3)=-2$ (based on the fact that it is non-compact
and that there is one complex deformation) we obtain
$$\int_{{\cal M}_g} c_{g-1}^3=C_{g,h\rightarrow 0}=(-1)^{g-1}\chi_g
{\zeta (2g-2)\over (2\pi)^{2g-2}}$$
in agreement with the mathematical result Eq.\cgeqn\ !

But can we also physically interpret the open string expansion ($h\neq 0$)
in terms of closed strings?  {\it Apriori}, we do not have any reason
to anticipate any relation
in this case.  However,
let us add the ``closed string sector'' ($p=0$) to the Chern-Simons
free energy and weigh terms with $2p$ holes with a factor of $(2\pi
A)^{2p}$.
In other words,(for $g >1$), we write the genus $g$ contribution
to the free energy as (remembering also the factor of $(-1)^{g-1+p}$)
$$F_g = {(-1)^{g-1}B_g\over 2g(2g-2)}\sum_{n=1}({\l\over 2\pi n})^{2g-2}
\sum_{p=0}(-1)^{g-1+p}{1\over n^{2p}}{\eqalign{\pmatrix{2g-3+2p\cr 2p}}}(2\pi
A)^{2p}$$
or on performing the sum over $p$,
$$F_g={(-1)^{g-1}B_g\over 2g(2g-2)}{\l^{2g-2}\over (2\pi)^{2g-2}}
\sum_{n=1}[{1\over (A+in)^{2g-2}}+
{1\over (A-in)^{2g-2}}].$$
This is what we had seen in the previous section as the contribution to
${\cal F}_g$ from membranes wrapped on $S^2$, with KK excitations of
momentum $n$.  It remains an open question to interpret
this result.  It is conceivable that the conifold transition from $S^3$ to
$S^2$,
which is geometrically understood,
is relevant here.

\newsec{Conclusions}
In this paper we showed how the BPS content of type IIA strings
on a Calabi-Yau threefold involving wrapped $D2$ branes and its bound
states with $D0$ branes (or equivalently $M2$ branes with Kaluza-Klein
momenta) can be used,
through a one loop computation
to reproduce all the $R^2 F^{2g-2}$ amplitudes that topological strings
compute.  In particular, a given BPS object affects all genus
$g$ topological amplitudes in a simple and computable way.  In this
paper, we exclusively dealt with the case of plain $D0$ branes
or their bound states
with $D2$ branes with the topology of (an isolated) $S^2$
in the Calabi-Yau.  This in particular allowed us to compute non-trivial
quantities over the moduli space of genus $g$ curves, as well as the
contribution of small worldsheet instantons in a physical way.
The fact that they agree with known/conjectured results
can be viewed as further confirmation of properties of D-branes
in type IIA context, or equivalently as further evidence
(or applications) of M-theory--type IIA duality. We also expect that
the non-perturbative form of the topological string partition function that
we arrived at, will yield future insights.

In a subsequent paper \nextp\ we extend these results to the
case where the $D2$ branes have higher genus (including also
wrapped multi D2 branes on continuous families of curves.)
The M-theory description
in conjunction with the type IIA description provides
a powerful way to compute these contributions, and thus allows
one to reformulate the entire topological string amplitudes in terms
of its BPS content.

It is natural to expect these results to also extend
to the $N=4$ topological strings \ref\berkv{N. Berkovits and C. Vafa, ``N=4
Topological Strings,'' Nucl. Phys. {\bf B433} (1995) 123.}.
  In particular the $R^4 F^{4g-4}$ amplitudes
of type IIA string in $R^{10}$ should have a similar $D0$ brane
interpretation \ref\bervv{N. Berkovits and C. Vafa,``Type IIB $R^4 H^{4g-4}$
Conjectures,'' hep-th/9803145.}.
  The same should be true for its compactification
on tori \ref\oov{H. Ooguri and C. Vafa, ``All Loop N=2 String Amplitudes,''
Nucl. Phys. {\bf B451} (1995) 121.}\
 or $K3$.  It would be extremely interesting
to relate the BPS content of these theories with the topological
$N=4$ amplitudes.

One possible application of our results is to questions involving 4 dimensional
black holes.  In particular the F-terms we have computed
correct the $R^2$ term in the effective 4 dimensional
action, in the presence of a non-vanishing
graviphoton
field strength.  Noting that the graviphoton field strength
is non-vanishing in a black hole background we conclude
that there are $R^2$ corrections which would affect the black
hole geometry.

\centerline{\bf Acknowledgments}

We would like to thank N. Berkovits, M. Bershadsky,
C. Faber, S. Katz, A. Lawrence,
J. Maldacena,
N. Nekrasov and A. Strominger for valuable discussions.

The research of R.G. is supported by DOE grant
DE-FG02-91 ER40654 and that of C.V. is supported in part by NSF grant
PHY-98-02709

\bigskip

\centerline{\bf Appendix}

Here we explicitly perform the 't Hooft expansion of the Chern-Simons free
energy on $S^3$. The theory is defined on a three manifold $M$ as
$$Z[M,N,k] = \int [DA] \exp{[{ik\over 4\pi}\int_M Tr(A\wedge dA+{2\over
3}A\wedge A\wedge A)]}.$$

On the sphere with gauge group $SU(N)$, the exact partition function is
\eqn\sphere{Z[S^3,N,k]= e^{{i\pi \over 8}N(N-1)}{1\over (N+k)^{N/2}}
\sqrt{N+k \over
N}\prod_{j=1}^{N-1}\{2\sin({j\pi\over N+k})\}^{N-j}.}
We'd like to examine the ${1\over N}$ expansion of the free energy in the
form Eqn.\t Hooft . The bare 't Hooft coupling in the Chern-Simons theory
is $\lambda_b = {2\pi N\over k}$. But it is known that there is a finite
renormalisation of the coupling such that $k \rightarrow k+N$. In other
words, the renormalised 't Hooft coupling $\lambda={2\pi N\over k+N}$. This is
the parameter we should be expanding in a perturbative (large $k$ )
expansion.

Therefore, as in \peri,
$$ F(S^3,N,\lambda)={-N\over 2}\ln N+{N-1\over 2}\ln \l
+\sum_{j=1}^{N-1}(N-j)[\ln j + \ln({\l\over
2\pi N})+\sum_{n=1}ln(1-{j^2\l^2\over 4\pi^2n^2N^2})]$$
(Using $sin(\pi x)=\pi x \prod_{n=1}(1-{x^2\over n^2})$)
We will focus on the last term in the sum, i.e.
$$\tilde{F}\equiv \sum_{j=1}^{N-1}(N-j)\sum_{n=1}ln(1-{j^2\l^2\over
4\pi^2n^2N^2}).$$
(The term in the $F(S^3, N,\l)$  given by $\sum_{j=1}^{N-1}(N-j)\ln j=\ln
G(N+1)$,
where the G-function is defined recursively as
$G(z+1)=\Gamma(z)G(z)$. This term is independent of $\l$ and will not
play a role in the following, though the term related to it by $N,k$
duality is the one that survives in the $N=\infty$ limit and is related to
IIB topological amplitudes \twobcs\ .)
Expanding the logarithm in the second term and carrying out the sum over
$n$ gives
$$\sum_{m=1}{\zeta(2m)\over m}({\l\over 2\pi
N})^{2m}\sum_{j=1}^{N-j}(N-j)j^{2m}.$$
The sum over $j$ can be carried out so that the
final answer is
$$\sum_{j=1}^{N-j}(N-j)j^{2m}={N^{2m+2}\over (2m+1)(2m+2)}+2m\sum_{g=1}^m
{\eqalign{\pmatrix{2m-1\cr 2g-3}}}(-1)^g {B_g\over 2g(2g-2)}N^{2m+2-2g}.$$
Then making the replacement for $g>0$, $(m=g-1+p)$, the relevant part of the
free
energy then reads as
$$\tilde{F}= \sum_{p=2}^{\infty}N^2{\zeta(2p-2)\over p-1}{({\l\over
2\pi})^{2p-2}
\over 2p(2p-1)} -\sum_{p=1}^{\infty}B_1{\zeta(2p)\over 2p}({\l\over
2\pi})^{2p}$$
$$\qquad + \sum_{g=2}^{\infty}N^{2-2g}{(-1)^gB_g\over
2g(2g-2)}\sum_{p=1}^{\infty}\zeta(2g-2+2p){\eqalign{\pmatrix{2g-3+2p\cr 2p}}}
({\l\over 2\pi})^{2g-2+2p}.$$

\listrefs

\end